\documentclass[useAMS,usenatbib,usegraphicx]{mn2e}

\usepackage{color}

\title [A structure at $z \sim 1.3$ that exceeds the homogeneity scale]
{A structure in the early universe at {$\bf z \sim 1.3$} that exceeds the
homogeneity scale of the R-W concordance cosmology}

\author[R.G. Clowes et al.]
{Roger G.~Clowes,$^1$\thanks{E-mail: rgclowes@uclan.ac.uk}
Kathryn A. Harris,$^1$
Srinivasan Raghunathan,$^{1,2}$\thanks{Present address: Universidad de Chile}
\newauthor
Luis E. Campusano,$^2$
Ilona K. S\"ochting$^3$
and
Matthew J. Graham$^4$ \\
$^1$ Jeremiah Horrocks Institute, University of Central Lancashire,
Preston PR1 2HE \\
$^2$ Observatorio Astron\'omico Cerro Cal\'an, Departamento de Astronom\'{\i}a,
Universidad de Chile, Casilla 36-D, Santiago, Chile \\
$^3$ Astrophysics, Denys Wilkinson Building, Keble Road,
University of Oxford, Oxford OX1 3RH \\
$^4$ California Institute of Technology, 1200 East California Boulevard,
Pasadena, CA 91125, USA}

\begin{document}

\date{Accepted 2012 November 24. Received 2012 November 12; in original form
2012 October 12}

\pagerange{\pageref{firstpage}--\pageref{lastpage}} \pubyear{2011}

\maketitle

\label{firstpage}

\begin{abstract}
A Large Quasar Group (LQG) of particularly large size and high membership has
been identified in the DR7QSO catalogue of the Sloan Digital Sky Survey. It
has characteristic size (volume$^{1/3}$) $\sim$ 500~Mpc (proper size, present
epoch), longest dimension $\sim$ 1240~Mpc, membership of 73 quasars, and mean
redshift $\bar{z} = 1.27$. In terms of both size and membership it is the
most extreme LQG found in the DR7QSO catalogue for the redshift range $1.0
\le z \le 1.8$ of our current investigation. Its location on the sky is $\sim
8.8^\circ$ north ($\sim$ 615~Mpc projected) of the Clowes \& Campusano LQG at
the same redshift, $\bar{z} = 1.28$, which is itself one of the more extreme
examples. Their boundaries approach to within $\sim 2^\circ$ ($\sim$ 140~Mpc
projected). This new, huge LQG appears to be the largest structure currently
known in the early universe. Its size suggests incompatibility with the 
Yadav et al. scale of homogeneity for the concordance cosmology, and thus
challenges the assumption of the cosmological principle.
\end{abstract}


\section{Introduction}

Large Quasar Groups (LQGs) are the largest structures seen in the early
universe, of characteristic size $\sim$ 70--350~Mpc, with the highest values
appearing to be only marginally compatible with the \citet*{Yadav2010} scale
of homogeneity in the concordance cosmology. LQGs generally have $\sim$ 5--40
member quasars. The first three LQGs to be discovered were those of:
\citet{Webster1982}; \citet*{Crampton1987}, \citet*{Crampton1989}; and
\citet{Clowes1991}. For more recent work see, for example: \citet{Brand2003}
(radio galaxies); \citet{Miller2004}; \citet{Pilipenko2007};
\citet{Rozgacheva2012}; and \citet{Clowes2012}. The association of quasars
with superclusters in the relatively local universe has been discussed by,
for example: \citet{Longo1991}; \citet*{Soechting2002};
\citet*{Soechting2004}; and \citet{Lietzen2009}. The last three of these
papers note the association of quasars with the peripheries of clusters or
with filaments. At higher redshifts, \citet*{Komberg1996} and
\citet{Pilipenko2007} suggest that the LQGs are the precursors of the
superclusters seen today. Given the large sizes of LQGs, perhaps they are
instead the precursors of supercluster complexes such as the Sloan Great Wall
\citep{Gott2005}.

In \citet{Clowes2012} we presented results for two LQGs as they appeared in
the DR7 quasar catalogue \citep[``DR7QSO'',][]{Schneider2010} of the Sloan
Digital Sky Survey (SDSS). One of these LQGs, designated U1.28 in that paper,
was the previously known \citet{Clowes1991} LQG (CCLQG) and the other,
designated U1.11, was a new discovery. (In these designations U1.28 and U1.11
the ``U'' refers to a connected unit of quasars, and the number refers to the
mean redshift.) U1.28 and U1.11 had memberships of 34 and 38 quasars
respectively, and characteristic sizes (volume$^{1/3}$) of $\sim 350,
380$~Mpc. \citet{Yadav2010} give an idealised upper limit to the scale of
homogeneity in the concordance cosmology as $\sim$ 370~Mpc. As discussed in
\citet{Clowes2012}, if the fractal calculations of \citet{Yadav2010} are
adopted as reference then U1.28 and U1.11 are only marginally compatible with
homogeneity.

In this paper we present results for a new LQG, designated U1.27, again found
in the DR7QSO catalogue, which is noteworthy for both its exceptionally large
characteristic size, $\sim$ 500~Mpc, and its exceptionally high membership,
73 quasars. It provides further interest for discussions of homogeneity and
the validity of the cosmological principle.

For simplicity we shall also refer to U1.27 as the Huge-LQG and U1.28 as the
CCLQG (for Clowes \& Campusano LQG).

The largest structure in the local universe is the Sloan Great Wall (SGW) at
$z = 0.073$, as noted in particular by \citet{Gott2005}. They give its length
(proper size at the present epoch) as $\sim$ 450~Mpc, compared with $\sim$
240~Mpc for the \citet{Geller1989} Great Wall ($z=0.029$).
Although \citet{Gott2005} do not discuss in detail the compatibility of the
SGW with concordance cosmology and gaussian initial conditions, from visual
inspection of simulations they did not expect any incompatibility.
\citet{Sheth2011} have investigated this question of compatibility further
and concluded that, given the assumptions of their analysis, there is a
potential difficulty, which can, however, be avoided if the SGW, in our
cosmological neighbourhood, happens to be the densest structure of its volume
within the entire Hubble volume.

The \citet{Sheth2011} paper is, however, not an analysis of compatibility of
the SGW with homogeneity.  Homogeneity asserts that the mass-energy density
(or, indeed, any global property) of sufficiently large volumes should be the
same within the expected statistical variations. \citet{Sheth2011} estimate
the volume of the SGW as $\sim 2.1 \times 10^6$~Mpc$^3$, for the larger of
two group-linkage estimates, which roughly reproduces the portrayal of the
SGW by \citet{Gott2005}. A characteristic size --- (volume)$^{1/3}$ --- is
then $\sim 128$~Mpc. The SGW is markedly elongated so this measure of
characteristic size should not be compared with the overall length. The
overdensity is given as $\delta_M \sim 1.2$ for mass and $\delta_n \sim 4$
for number of galaxies. Note that \citet{Einasto2011c} find that the SGW is
not a single structure, but a set of superclusters with different
evolutionary histories. For discussing potential conflicts of the SGW with
homogeneity this result by \citet{Einasto2011c} means that the long dimension
of $\sim$ 450~Mpc is misleading. The characteristic size of $\sim 128$~Mpc is
still relevant, but is much smaller than the upper limit of $\sim 370$~Mpc
for homogeneity \citep{Yadav2010}, and so it may be that the SGW does not
present any particular problem. Indeed, \citet{Park2012} find from the
``Horizon Run 2'' cosmological simulation that the SGW is consistent with
concordance cosmology and with homogeneity. \citet{Park2012} also note that
the properties of LSSs can be used as sensitive discriminants of cosmological
models and models of galaxy formation.

The concordance model is adopted for cosmological calculations, with
$\Omega_T = 1$, $\Omega_M = 0.27$, $\Omega_\Lambda = 0.73$, and $H_0 =
70$~kms$^{-1}$Mpc$^{-1}$. All sizes given are proper sizes at the present
epoch.

\section{Detection of the Huge-LQG (U1.27)}

The new, Huge-LQG (U1.27) has been detected by the procedures described in
\citet{Clowes2012}. These procedures are briefly described here.

As mentioned above, the source of the quasar data is the SDSS DR7QSO
catalogue \citep{Schneider2010} of 105783 quasars. The low-redshift, $z \la
3$, strand of selection of the SDSS specifies $i \le 19.1$
\citep{Richards2006, Vanden-Berk2005}. Restriction of the quasars to this
limit allows satisfactory spatial uniformity of selection on the sky to be
achieved, since they are then predominantly from this strand. Also, changes
in the SDSS selection algorithms \citep{Richards2002} should not then be
important. The more general criteria for extraction of a statistical sample
from the DR7QSO catalogue or its predecessors are discussed by
\citet{Schneider2010, Richards2006, Vanden-Berk2005}.

The DR7QSO catalogue covers $\sim$ 9380~deg$^2$ in total. There is a large
contiguous area of $\sim$ 7600~deg$^2$ in the north galactic gap, which has
some jagged boundaries. Within this contiguous area we define a control area,
designated A3725, of $\sim$ 3725 deg$^2$ (actually 3724.5 deg$^2$) by RA:
$123.0^\circ \rightarrow 237.0^\circ$ and Dec: $15.0^\circ \rightarrow
56.0^\circ$.

We detect candidates for LQGs in the catalogue by three-dimensional
single-linkage hierarchical clustering, which is equivalent to the
three-dimensional minimal spanning tree (MST). Such algorithms have the
advantage that they do not require assumptions about the morphology of the
structure. As in \citet{Clowes2012}, the linkage scale is set to 100~Mpc. The
choice of scale is guided by the mean nearest-neighbour separation together
with allowance for redshift errors and peculiar velocities; see that paper
for full details. The particular algorithm we use for single-linkage
hierarchical clustering is the {\it agnes\/} algorithm in the {\it R
  package}\footnote{See http://www.r-project.org}. We are currently
concentrating on detecting LQGs in the redshift interval $1.0 \le z \le 1.8$
and, of course, with the restriction $i \le 19.1$.

With this detection procedure the new Huge-LQG (U1.27) that is the subject of
this paper is detected as a unit of 73 quasars, with mean redshift 1.27. It
covers the redshift range $1.1742 \rightarrow 1.3713$. The 73 member quasars
are listed in Table~\ref{Huge-LQG_table}.

The Huge-LQG is $\sim 8.8^\circ$ north ($\sim$ 615~Mpc projected) of the
CCLQG at the same redshift. Their boundaries on the sky approach to within
$\sim 2^\circ$ ($\sim$ 140~Mpc projected).

\begin {table*}
\flushleft
\caption {Huge-LQG (U1.27): the set of 73 100Mpc-linked quasars from the SDSS
DR7QSO catalogue. The columns are: SDSS name; RA, Dec. (2000); redshift;
$i$ magnitude.}
\small \renewcommand \arraystretch {0.8}
\newdimen\padwidth
\setbox0=\hbox{\rm0}
\padwidth=0.3\wd0
\catcode`|=\active
\def|{\kern\padwidth}
\newdimen\digitwidth
\setbox0=\hbox{\rm0}
\digitwidth=0.7\wd0
\catcode`!=\active
\def!{\kern\digitwidth}
\begin {tabular} {lllll}
\\
SDSS name            & RA, Dec (2000)             & z      & $i $              \\
                     &                            &        &                   \\
\\
104139.15$+$143530.2 & 10:41:39.15~~$+$14:35:30.2 & 1.2164 & 18.657            \\
104321.62$+$143600.2 & 10:43:21.62~~$+$14:36:00.2 & 1.2660 & 19.080            \\
104430.92$+$160245.0 & 10:44:30.92~~$+$16:02:45.0 & 1.2294 & 17.754            \\
104445.03$+$151901.6 & 10:44:45.03~~$+$15:19:01.6 & 1.2336 & 18.678            \\
104520.62$+$141724.2 & 10:45:20.62~~$+$14:17:24.2 & 1.2650 & 18.271            \\
104604.05$+$140241.2 & 10:46:04.05~~$+$14:02:41.2 & 1.2884 & 18.553            \\
104616.31$+$164512.6 & 10:46:16.31~~$+$16:45:12.6 & 1.2815 & 18.732            \\
104624.25$+$143009.1 & 10:46:24.25~~$+$14:30:09.1 & 1.3620 & 18.989            \\
104813.63$+$162849.1 & 10:48:13.63~~$+$16:28:49.1 & 1.2905 & 18.593            \\
104859.74$+$125322.3 & 10:48:59.74~~$+$12:53:22.3 & 1.3597 & 18.938            \\
104915.66$+$165217.4 & 10:49:15.66~~$+$16:52:17.4 & 1.3459 & 18.281            \\
104922.60$+$154336.1 & 10:49:22.60~~$+$15:43:36.1 & 1.2590 & 18.395            \\
104924.30$+$154156.0 & 10:49:24.30~~$+$15:41:56.0 & 1.2965 & 18.537            \\
104938.22$+$214829.3 & 10:49:38.22~~$+$21:48:29.3 & 1.2352 & 18.805            \\
104941.67$+$151824.6 & 10:49:41.67~~$+$15:18:24.6 & 1.3390 & 18.792            \\
104947.77$+$162216.6 & 10:49:47.77~~$+$16:22:16.6 & 1.2966 & 18.568            \\
104954.70$+$160042.3 & 10:49:54.70~~$+$16:00:42.3 & 1.3373 & 18.748            \\
105001.22$+$153354.0 & 10:50:01.22~~$+$15:33:54.0 & 1.2500 & 18.740            \\
105042.26$+$160056.0 & 10:50:42.26~~$+$16:00:56.0 & 1.2591 & 18.036            \\
105104.16$+$161900.9 & 10:51:04.16~~$+$16:19:00.9 & 1.2502 & 18.187            \\
105117.00$+$131136.0 & 10:51:17.00~~$+$13:11:36.0 & 1.3346 & 19.027            \\
105119.60$+$142611.4 & 10:51:19.60~~$+$14:26:11.4 & 1.3093 & 19.002            \\
105122.98$+$115852.3 & 10:51:22.98~~$+$11:58:52.3 & 1.3085 & 18.127            \\
105125.72$+$124746.3 & 10:51:25.72~~$+$12:47:46.3 & 1.2810 & 17.519            \\
105132.22$+$145615.1 & 10:51:32.22~~$+$14:56:15.1 & 1.3607 & 18.239            \\
105140.40$+$203921.1 & 10:51:40.40~~$+$20:39:21.1 & 1.1742 & 17.568            \\
105144.88$+$125828.9 & 10:51:44.88~~$+$12:58:28.9 & 1.3153 & 19.021            \\
105210.02$+$165543.7 & 10:52:10.02~~$+$16:55:43.7 & 1.3369 & 16.430            \\
105222.13$+$123054.1 & 10:52:22.13~~$+$12:30:54.1 & 1.3162 & 18.894            \\
105223.68$+$140525.6 & 10:52:23.68~~$+$14:05:25.6 & 1.2483 & 18.640            \\
105224.08$+$204634.1 & 10:52:24.08~~$+$20:46:34.1 & 1.2032 & 18.593            \\
105245.80$+$134057.4 & 10:52:45.80~~$+$13:40:57.4 & 1.3544 & 18.211            \\
105257.17$+$105933.5 & 10:52:57.17~~$+$10:59:33.5 & 1.2649 & 19.056            \\
105258.16$+$201705.4 & 10:52:58.16~~$+$20:17:05.4 & 1.2526 & 18.911            \\
105412.67$+$145735.2 & 10:54:12.67~~$+$14:57:35.2 & 1.2277 & 18.767            \\
105421.90$+$212131.2 & 10:54:21.90~~$+$21:21:31.2 & 1.2573 & 17.756            \\
105435.64$+$101816.3 & 10:54:35.64~~$+$10:18:16.3 & 1.2600 & 17.951            \\
105442.71$+$104320.6 & 10:54:42.71~~$+$10:43:20.6 & 1.3348 & 18.844            \\
105446.73$+$195710.5 & 10:54:46.73~~$+$19:57:10.5 & 1.2195 & 18.759            \\
105523.03$+$130610.7 & 10:55:23.03~~$+$13:06:10.7 & 1.3570 & 18.853            \\
105525.18$+$191756.3 & 10:55:25.18~~$+$19:17:56.3 & 1.2005 & 18.833            \\
105525.68$+$113703.0 & 10:55:25.68~~$+$11:37:03.0 & 1.2893 & 18.264            \\
105541.83$+$111754.2 & 10:55:41.83~~$+$11:17:54.2 & 1.3298 & 18.996            \\
105556.22$+$184718.4 & 10:55:56.22~~$+$18:47:18.4 & 1.2767 & 18.956            \\
105611.27$+$170827.5 & 10:56:11.27~~$+$17:08:27.5 & 1.3316 & 17.698            \\
105621.90$+$143401.0 & 10:56:21.90~~$+$14:34:01.0 & 1.2333 & 19.052            \\
105637.49$+$150047.5 & 10:56:37.49~~$+$15:00:47.5 & 1.3713 & 19.041            \\
105637.98$+$100307.2 & 10:56:37.98~~$+$10:03:07.2 & 1.2730 & 18.686            \\
105655.36$+$144946.2 & 10:56:55.36~~$+$14:49:46.2 & 1.2283 & 18.590            \\
105714.02$+$184753.3 & 10:57:14.02~~$+$18:47:53.3 & 1.2852 & 18.699            \\
105805.09$+$200341.0 & 10:58:05.09~~$+$20:03:41.0 & 1.2731 & 17.660            \\
105832.01$+$170456.0 & 10:58:32.01~~$+$17:04:56.0 & 1.2813 & 18.299            \\
105840.49$+$175415.5 & 10:58:40.49~~$+$17:54:15.5 & 1.2687 & 18.955            \\
105855.33$+$081350.7 & 10:58:55.33~~$+$08:13:50.7 & 1.2450 & 17.926            \\
105928.57$+$164657.9 & 10:59:28.57~~$+$16:46:57.9 & 1.2993 & 19.010            \\
110006.02$+$092638.7 & 11:00:06.02~~$+$09:26:38.7 & 1.2485 & 18.055            \\
110016.88$+$193624.7 & 11:00:16.88~~$+$19:36:24.7 & 1.2399 & 18.605            \\
110039.99$+$165710.3 & 11:00:39.99~~$+$16:57:10.3 & 1.2997 & 18.126            \\
110148.66$+$082207.1 & 11:01:48.66~~$+$08:22:07.1 & 1.1940 & 18.880            \\
110217.19$+$083921.1 & 11:02:17.19~~$+$08:39:21.1 & 1.2355 & 18.800            \\
110504.46$+$084535.3 & 11:05:04.46~~$+$08:45:35.3 & 1.2371 & 19.005            \\
110621.40$+$084111.2 & 11:06:21.40~~$+$08:41:11.2 & 1.2346 & 18.649            \\
110736.60$+$090114.7 & 11:07:36.60~~$+$09:01:14.7 & 1.2266 & 18.902            \\
110744.61$+$095526.9 & 11:07:44.61~~$+$09:55:26.9 & 1.2228 & 17.635            \\
111007.89$+$104810.3 & 11:10:07.89~~$+$10:48:10.3 & 1.2097 & 18.473            \\
111009.58$+$075206.8 & 11:10:09.58~~$+$07:52:06.8 & 1.2123 & 18.932            \\
111416.17$+$102327.5 & 11:14:16.17~~$+$10:23:27.5 & 1.2053 & 18.026            \\
111545.30$+$081459.8 & 11:15:45.30~~$+$08:14:59.8 & 1.1927 & 18.339            \\
111802.11$+$103302.4 & 11:18:02.11~~$+$10:33:02.4 & 1.2151 & 17.486            \\
111823.21$+$090504.9 & 11:18:23.21~~$+$09:05:04.9 & 1.1923 & 18.940            \\
112019.62$+$085905.1 & 11:20:19.62~~$+$08:59:05.1 & 1.2239 & 18.093            \\
112059.27$+$101109.2 & 11:20:59.27~~$+$10:11:09.2 & 1.2103 & 18.770            \\
112109.76$+$075958.6 & 11:21:09.76~~$+$07:59:58.6 & 1.2369 & 18.258            \\
\\
\end {tabular}
\label{Huge-LQG_table}
\end {table*}

\section{Properties of the Huge-LQG}

Groups found by the linkage of points require a procedure to assess their
statistical significance and to estimate the overdensity. We use the CHMS
method (``convex hull of member spheres''), which is described in detail by
\citet{Clowes2012}. The essential statistic is the volume of the candidate: a
LQG must occupy a smaller volume than the expectation for the same number of
random points.

In the CHMS method the volume is constructed as follows. Each member point of
a unit is expanded to a sphere, with radius set to be half of the mean
linkage (MST edge length) of the unit. The CHMS volume is then taken to be
the volume of the convex hull of these spheres. The significance of a LQG
candidate of membership $N$ is found from the distribution of CHMS volumes
resulting from 1000 sets of $N$ random points that have been distributed in a
cube of volume such that the density in the cube corresponds to the density
in a control area for the redshift limits of the candidate. The CHMS volumes
for the random sets can also be used to estimate residual biases
\citep[see][]{Clowes2012} and consequently make corrections to the properties
of the LQGs.

In this way, with A3725 as the control area, we find that the departure from
random expectations for the Huge-LQG is $3.81\sigma$. After correcting the
CHMS volumes for residual bias the estimated overdensity of the Huge-LQG is
$\delta_q = \delta \rho_q / \rho_q = 0.40$. (The volume correction is $\sim$
2 per cent.) The overdensity is discussed further below, because of
the cautious, conservative nature of the CHMS estimate, which, in this case,
is possibly too cautious.


As discussed in \citet{Clowes2012} a simple measure of the characteristic
size of an LQG, which takes no account of morphology, is the cube root of the
corrected CHMS volume. For the Huge-LQG the volume is $\sim 1.21 \times
10^8$~Mpc$^3$, giving a characteristic size of $\sim 495$~Mpc.

From the inertia tensor of the member quasars of the Huge-LQG, the principal
axes have lengths of $\sim$ 1240, 640, and 370 Mpc, and the inhomogeneity
thus extends to the Gpc-scale. The axis ratios are 3.32~:~1.71~:~1, so it is
substantially elongated.

\begin{figure*}
\includegraphics{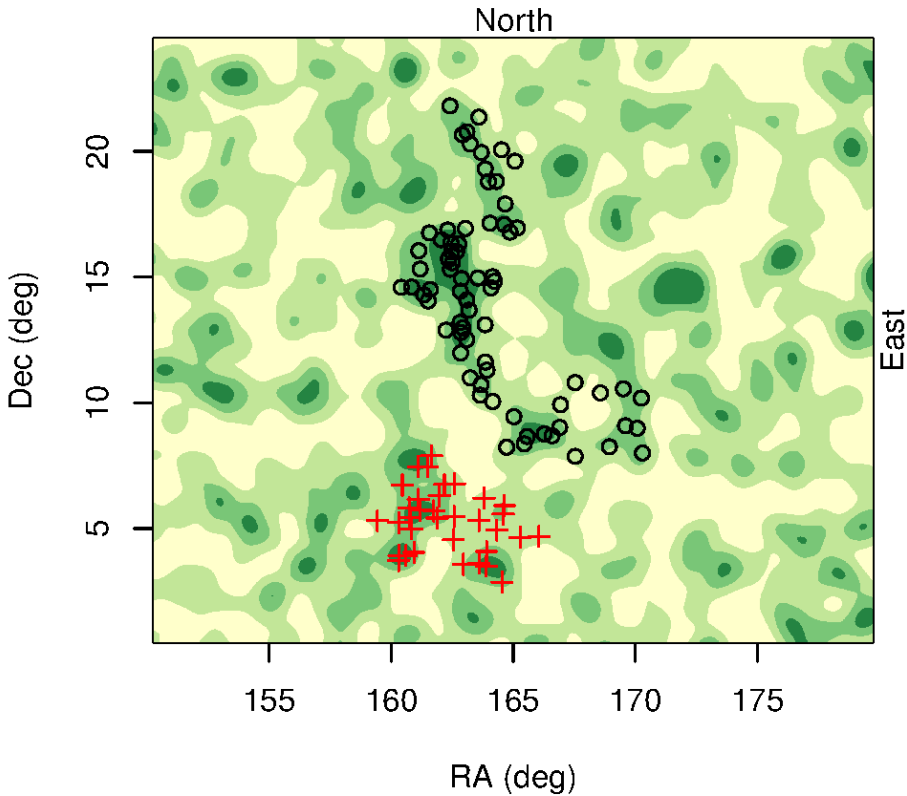}
\caption{The sky distribution of the 73 quasars of the new, Huge-LQG (U1.27,
$\bar{z}=1.27$, circles) is shown, together with that of the 34 quasars
of the \citet{Clowes1991} LQG, CCLQG ($\bar{z} = 1.28$, crosses). The
members of each LQG are connected at the linkage scale of 100~Mpc. The area
shown is approximately $29.5^\circ \times 24.0^\circ$. The DR7QSO quasars
are limited to $i \le 19.1$. Superimposed on these distributions is a
kernel-smoothed intensity map (isotropic Gaussian kernel, $\sigma =
0.5^\circ$), plotted with four linear palette levels ($\le 0.8$, 0.8--1.6,
1.6--2.4, $\ge 2.4$ deg$^{-2}$), for all of the quasars in the joint
redshift range of Huge-LQG and CCLQG ($z: 1.1742 \rightarrow 1.4232$). No
$\cos\delta$ correction has been applied to this intensity map.}
\label{skydist_Huge-LQG_CCLQG}
\end{figure*}

Fig.~\ref{skydist_Huge-LQG_CCLQG} shows the sky distributions of the members
of both the new, Huge-LQG (U1.27) and the Clowes \& Campusano LQG,
CCLQG. Much of the Huge-LQG is directly north of the CCLQG, but the southern
part curves to the south-east and away from the CCLQG. The redshift intervals
occupied by the two LQGs are similar (1.1742--1.3713 for the Huge-LQG,
1.1865--1.4232 for the CCLQG), but on the sky Huge-LQG is clearly
substantially larger.

\begin{figure*}
\includegraphics[height=120mm]{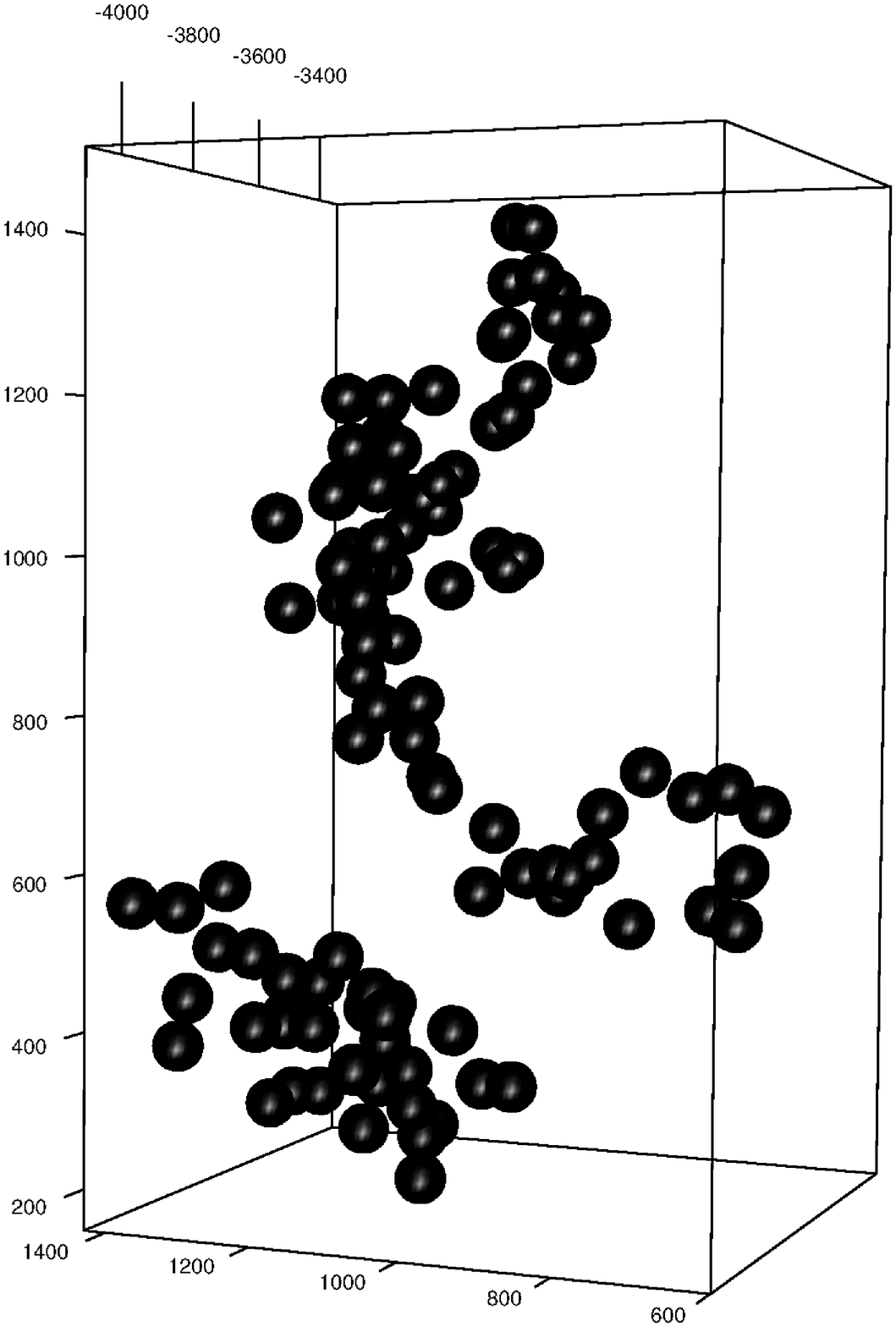}
\caption{A snapshot from a visualisation of both the new, Huge-LQG and the
CCLQG (the Clowes \& Campusano LQG). The scales shown on the cuboid are
proper sizes (Mpc) at the present epoch. The tick marks represent intervals
of 200~Mpc. The Huge-LQG appears as the upper LQG. For comparison, the
members of both are shown as spheres of radius 33.0~Mpc (half of the mean
linkage for the Huge-LQG; the value for the CCLQG is 38.8~Mpc). For the
Huge-LQG, note the dense, clumpy part followed by a change in orientation
and a more filamentary part. The Huge-LQG and the CCLQG appear to be
distinct entities.}
\label{visualisation_Huge-LQG_CCLQG}
\end{figure*}

Fig.~\ref{visualisation_Huge-LQG_CCLQG} shows a snapshot from a visualisation
of the new, Huge-LQG, and CCLQG, the Clowes \& Campusano LQG. The scales
shown on the cuboid are proper sizes (Mpc) at the present epoch. The member
points of both LQGs are shown expanded to spheres of radius 33.0~Mpc, which
is half of the mean linkage (MST edge length) for the Huge-LQG (consistent
with the CHMS method for this LQG). The morphology of the Huge-LQG is clearly
strongly elongated, and curved. There is the appearance of a dense, clumpy
part, followed by a change in orientation and a more filamentary part. Note
that half of the mean linkage for the CCLQG is actually 38.8~Mpc, so, in this
respect, the Huge-LQG is more tightly connected than the CCLQG. However, the
CHMS-density is lower for Huge-LQG than for CCLQG because of the effect of
the change in orientation on the CHMS of Huge-LQG. That is, the Huge-LQG is
more tightly connected than CCLQG (33.0~Mpc compared with 38.8~Mpc) but its
curvature causes its CHMS-volume to be disproportionately large (there is
more ``dead space'') and hence its density to be disproportionately low.
Note that the Huge-LQG and the CCLQG appear to be distinct entities --- their
CHMS volumes do not intersect.

The CHMS method is thus conservative in its estimation of volume and hence of
significance and overdensity. Curvature of the structure can lead to the CHMS
volume being substantially larger than if it was linear. If we divide the
Huge-LQG into two sections at the point at which the direction appears to
change then we have a ``main'' set of 56 quasars and a ``branch'' set of 17
quasars. If we calculate the CHMS volumes of the main set and the branch
set, using the same sphere radius (33~Mpc) as for the full set of 73, and
simply add them (neglecting any overlap), then we obtain $\delta_q = \delta
\rho_q / \rho_q = 1.12$, using the same correction for residual bias (2 per
cent) as for the full set. That is, we have calculated $\delta_q$ using the
total membership (73) and the summed volume of the main set and the branch
set, and the result is now $\delta_q \sim 1$, rather than $\delta_q = 0.40$,
since much of the ``dead space'' has been removed from the volume estimate.

We should consider the possibility that the change in direction is indicating
that, physically if not algorithmically, we have two distinct structures at
the same redshift. So, if we instead treat the main and branch sets as two
independent LQG candidates and use their respective sphere radii for
calculation of CHMS volumes, including their respective corrections for
residual bias, then we obtain the following parameters. Main set of 56:
significance $5.86\sigma$; $\delta_q = 1.20$; characteristic size
(CHMS-volume$^{1/3}$) 390~Mpc; mean linkage 65.1~Mpc; and principal axes of
the inertia tensor $\sim$ 930, 410, 320~Mpc. Branch set of 17: significance
$2.91\sigma$; $\delta_q = 1.54$; characteristic size (CHMS-volume$^{1/3}$)
242~Mpc; mean linkage 67.7~Mpc; and principal axes of the inertia tensor
$\sim$ 570, 260, 150~Mpc. The similarity of the mean linkages suggests, after
all, a single structure with curvature rather than two distinct structures.
(Note for comparison that the CCLQG has mean linkage of 77.5~Mpc.) A
two-sided Mann-Whitney test finds no significant differences of the linkages
for the main and branch sets, which again suggests a single structure. Note
also that the main set by itself exceeds the \citet{Yadav2010} scale of
homogeneity.

We can estimate the masses of these main and branch sets from their CHMS
volumes by assuming that $\delta_q \equiv \delta_M$, where $\delta_M$ refers
to the mass in baryons and dark matter ($\Omega_M = 0.27$). We find that the
mass contained within the main set is $\sim 4.8 \times 10^{18} M_\odot$ and
within the branch set is $\sim 1.3 \times 10^{18} M_\odot$. Compared with the
expectations for their volumes these values correspond to mass excesses of
$\sim 2.6 \times 10^{18} M_\odot$ and $\sim 0.8 \times 10^{18} M_\odot$
respectively. The total mass excess is then $\sim 3.4 \times 10^{18}
M_\odot$, equivalent to $\sim$ 1300 Coma clusters \citep{Kubo2007}, $\sim$ 50
Shapley superclusters \citep{Proust2006}, or $\sim 20$ Sloan Great Walls
\citep{Sheth2011}.

\subsection{Corroboration of the Huge-LQG from Mg{\sc II} absorbers}

\begin{figure*}
\includegraphics{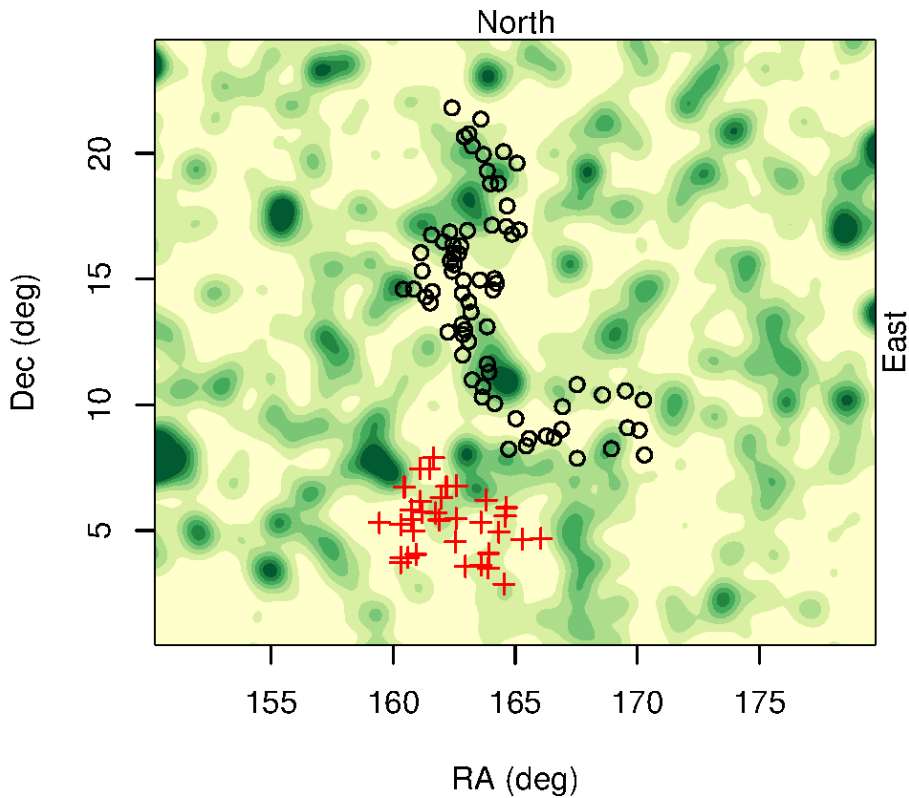}
\caption{The sky distribution of the 73 quasars of the new, Huge-LQG
($\bar{z}=1.27$, circles) is shown, together with that of the 34 quasars
of the \citet{Clowes1991} LQG, CCLQG ($\bar{z} = 1.28$, crosses). The
members of each LQG are connected at the linkage scale of 100~Mpc. The area
shown is approximately $29.5^\circ \times 24.0^\circ$. The DR7QSO quasars
are limited to $i \le 19.1$ for the LQG members. Superimposed on these
distributions is a kernel-smoothed intensity map (isotropic Gaussian
kernel, $\sigma = 0.5^\circ$), plotted with seven linear palette levels
($\le 0.62$, 0.62--1.24, 1.24--1.86, 1.86--2.48, 2.48--3.10, 3.10--3.72,
$\ge 3.72$ deg$^{-2}$), for all of the Mg{\sc II} $\lambda\lambda
2796,2798$ absorbers in the joint redshift range of the Huge-LQG and the
CCLQG ($z: 1.1742 \rightarrow 1.4232$) that have been found in the DR7QSO
background quasars ($z > 1.4232$, non-BAL, and restricted to $i \le 19.1$)
using the Mg{\sc II} absorber catalogue of \citet{Raghunathan2012}. The
Mg{\sc II} systems used here have rest-frame equivalent widths for the
$\lambda 2796$ component of $0.5 \le W_{r,2796} \le 4.0 \AA$. Apparent
Mg{\sc II} systems occurring shortward of the Ly-$\alpha$ emission in the
background quasars are assumed to be spurious and have been excluded. No
$\cos\delta$ correction has been applied to this intensity map.}
\label{mgii_Huge-LQG_CCLQG}
\end{figure*}

Some independent corroboration of this large structure is provided by Mg{\sc
  II} absorbers. We have used the DR7QSO quasars in a survey for intervening
Mg{\sc II} $\lambda\lambda 2796,2798$ absorbers
\citep{Raghunathan2012}. Using this survey, Fig.~\ref{mgii_Huge-LQG_CCLQG}
shows a kernel-smoothed intensity map (similar to
Fig.~\ref{skydist_Huge-LQG_CCLQG}) of the Mg{\sc II} absorbers across the
field of the Huge-LQG and the CCLQG, and for their joint redshift range ($z:
1.1742 \rightarrow 1.4232$). For this map, only DR7QSO quasars with $z >
1.4232$ have been used as probes of the Mg{\sc II} --- that is, only quasars
beyond the LQGs, and none within them. However, background quasars that are
known from the DR7QSO ``Catalog of Properties'' \citep{Shen2011} to be BAL
(broad absorption line) quasars have been excluded because structure within
the BAL troughs can lead to spurious detections of MgII doublets at similar
apparent redshifts. The background quasars have been further restricted to $i
\le 19.1$ for uniformity of coverage. A similar kernel-smoothed intensity map
(not shown here) verifies that the distribution of the used background
quasars is indeed appropriately uniform across the area of the figure.

The Mg{\sc II} systems used here have rest-frame equivalent widths for the
$\lambda 2796$ component of $0.5 \le W_{r,2796} \le 4.0 \AA$. For the
resolution and signal-to-noise ratios of the SDSS spectra, this lower limit
of $W_{r,2796} = 0.5 \AA$ appears to give consistently reliable detections,
although, being ``moderately strong'', it is higher than the value of
$W_{r,2796} = 0.3 \AA$ that would typically be used with spectra from larger
telescopes. Note that apparent Mg{\sc II} systems occurring shortward of the
Ly-$\alpha$ emission in the background quasars are assumed to be spurious and
have been excluded.

The RA-Dec track of the Huge-LQG quasars, along the $\sim 12^\circ$ where the
surface density is highest, appears to be closely associated with the track
of the Mg{\sc II} absorbers. The association becomes a little weaker in the
following $\sim 5^\circ$, following the change in direction from the main set
to the branch set, where the surface density of the quasars becomes
lower. Note that the quasars tend to follow the periphery of the structure in
the Mg{\sc II} absorbers, which is reminiscent of the finding by
\citet{Soechting2002} and \citet{Soechting2004} that quasars tend to lie on
the peripheries of galaxy clusters.

Note that the CCLQG is less clearly detected in Mg{\sc II} here, although it
was detected by \citet{Williger2002}. \citet{Williger2002} were able to
achieve a lower equivalent-width limit $W_{r,2796} = 0.3 \AA$ with their
observations on a 4m telescope. Furthermore, Fig.~\ref{mgii_Huge-LQG_CCLQG}
shows that the surface density of the Huge-LQG quasars is clearly higher than
for the CCLQG quasars, which is presumably a factor in the successful
detection of corresponding Mg{\sc II} absorption. The high surface density of
the members of the Huge-LQG seems likely to correspond to a higher
probability of lines of sight to the background quasars intersecting the
haloes of galaxies at small impact parameters.

\section{Discussion of homogeneity, and conclusions}

In \citet{Clowes2012} we presented results for the CCLQG, the Clowes \&
Campusano LQG, as it appeared in the SDSS DR7QSO catalogue, and also for
U1.11, a newly-discovered LQG in the same cosmological neighbourhood. We
noted that their characteristic sizes, defined as (CHMS-volume)$^{1/3}$, of
$\sim$ 350 and 380~Mpc respectively were only marginally compatible with the
\citet{Yadav2010} 370-Mpc upper limit to the scale of homogeneity for the
concordance cosmology. Their long dimensions from the inertia tensor of
$\sim$ 630 and 780~Mpc are clearly much larger.

In the present paper we have presented results for the Huge-LQG, another
newly discovered LQG from the DR7QSO catalogue, that is at essentially the
same redshift as the CCLQG, and only a few degrees to the north of it. It has
73 member quasars, compared with 34 and 38 for the CCLQG and U1.11. MgII
absorbers in background quasars provide independent corroboration of this
extraordinary LQG. The characteristic size of (CHMS-volume)$^{1/3}$ $\sim$
495~Mpc is well in excess of the \citet{Yadav2010} homogeneity scale, and the
long dimension from the inertia tensor of $\sim$ 1240~Mpc is spectacularly
so. It appears to be the largest feature so far seen in the early
universe. Even the ``main'' set alone, before the change of direction leading
to the ``branch'' set, exceeds the homogeneity scale. This Huge-LQG thus
challenges the assumption of the cosmological principle.  Its excess mass,
compared with expectations for its (main $+$ branch) volume, is $\sim 3.4
\times 10^{18} M_\odot$, equivalent to $\sim$ 1300 Coma clusters,
$\sim$ 50 Shapley superclusters, or $\sim 20$ Sloan Great Walls.

The usual models of the universe in cosmology, varying only according to the
parameter settings, are built on the assumption of the cosmological principle
- that is, on the assumption of homogeneity after imagined smoothing on some
suitably large scale. In particular, the models depend on the
Robertson-Walker metric, which assumes the homogeneity of the mass-energy
density. Given the further, sensible assumption that any property of the
universe ultimately depends on the mass-energy content then homogeneity
naturally asserts that any global property of sufficiently large volumes
should be the same within the expected statistical variations. A recent
review of {\it inhomogeneous\/} models is given by \citet{Buchert2011}.

We adopt the \citet{Yadav2010} fractal calculations as our reference for the
upper limit to the scale of homogeneity in the concordance model of
cosmology: inhomogeneities should not be detectable above this limit of
$\sim$ 370~Mpc. The \citet{Yadav2010} calculations have the appealing
features that the scale of homogeneity is essentially independent of both the
epoch and the tracer used. Note that the scale of $\sim$ 370~Mpc is much
larger than the scales of $\sim$ 100--115~Mpc for homogeneity deduced by
\citet{Scrimgeour2012}, and, for our purposes, it is therefore appropriately
cautious.

The cosmic microwave background is usually considered to provide the best
evidence for isotropy, and hence of homogeneity too, given the assumption of
isotropy about all points. Nevertheless, there do appear to be large-scale
features in the CMB that may challenge the reality of homogeneity and
isotropy --- see \citet{Copi2010} for a recent review. More recently still
than this review, \citet{Rossmanith2012} find further indications of a
violation of statistical isotropy in the CMB. Furthermore,
\citet*{Yershov2012} find that supernovae in the redshift range 0.5--1.0 are
associated with systematic CMB temperature fluctuations, possibly arising
from large-scale inhomogeneities. Observationally, for SDSS DR7 galaxies with
$0.22 < z < 0.50$, \citet*{Marinoni2012} find that isotropy about all points
does indeed apply on scales larger than $\sim$ 210~Mpc.

While \citet{Scrimgeour2012} find a transition to homogeneity on scales
$\sim$ 100--115~Mpc, using WiggleZ data, \citet{SylosLabini2011} does not, on
scales up to $\sim$ 200~Mpc, using SDSS galaxies. Large inhomogeneities in
the distribution of superclusters (supercluster complexes) such as the Sloan
Great Wall and in the voids have also been found on scales $\sim$
200--300~Mpc by \citet{Einasto2011b}, \citet*{Liivamagi2012},
\citet{Luparello2011}, and earlier references given within these
papers. Evidence for Gpc-scale correlations of galaxies has been presented
by, for example, \citet{Nabokov2008}, \citet{Padmanabhan2007} and
\citet*{Thomas2011}. The occurrence of structure on Gpc-scales from the
Huge-LQG and from galaxies implies that the universe is not homogeneous on
these scales. Furthermore, if we accept that homogeneity refers to any
property of the universe then an intriguing result is that of
\citet{Hutsemekers2005}, who found that the polarisation vectors of quasars
are correlated on Gpc-scales. Similarly, the existence of cosmic flows on
approximately Gpc-scales \citep[e.g. ][]{Kashlinsky2010}, regardless of their
cause, is itself implying that the universe is not homogeneous.

Of course, history and, most recently, the work of \citet{Park2012} indicate
that one should certainly be cautious on the question of homogeneity and the
cosmological principle. The Sloan Great Wall \citep{Gott2005} --- and before
it, the Great Wall \citep{Geller1989} --- was seen as a challenge to the
standard cosmology and yet \citet{Park2012} show that, in the ``Horizon Run
2'' concordance simulation of box-side 10~Gpc, comparable and even larger
features can arise, although they are of course rare. Nevertheless, the
Huge-LQG presented here is much larger, and it is adjacent to the CCLQG,
which is itself very large, so the challenges still persist.

\citet{Park2012} find that void complexes on scales up to $\sim$ 450 Mpc are
also compatible with the concordance cosmology, according to their
simulations, although the scales here are greater than the \citet{Yadav2010}
scale of homogeneity and much greater than the \citet{Scrimgeour2012}
scale. Also, \citet{Frith2003} find evidence for a local void on scales
$\sim$ 430~Mpc. The question of what exactly is a ``void complex'' might need
further attention. It seems likely to correspond to the ``supervoid'' of
\citet{Einasto2011a} and earlier references given there.

\citet{Hoyle2012} have investigated homogeneity within the past light-cone,
rather than on it, using the fossil record of star formation and find no
marked variation on a scale of $\sim$ 340~Mpc for $0.025 < z < 0.55$

\citet{Jackson2012} finds, from ultra-compact radio sources limited to $z >
0.5$, that the universe is not homogeneous on the largest scales: there is
more dark matter in some directions than in others.

The Huge-LQG and the CCLQG separately and together would also indicate that
there is more dark matter in some directions than in others. Such mass
concentrations could conceivably be associated with the cosmic (dark) flows
on scales of $\sim$ 100-1000~Mpc as reported by, for example,
\citet{Kashlinsky2008}; \citet*{Watkins2009}; \citet*{Feldman2010}; and
\citet{Kashlinsky2010}. Of particular interest is the possibility raised by
\citet{Tsagas2012} that those living within a large-scale cosmic flow could
see local acceleration of the expansion within a universe that is
decelerating overall. Tsagas notes that the proximity of the supernova dipole
to the CMB dipole could support such an origin for the apparent acceleration
that we see. With quasars mostly extinguished by the present epoch, we would
probably have some difficulty in recognising the counterparts today of such
LQGs then that might cause such cosmic flows. Very massive structures in the
relatively local universe could conceivably be present, but unrecognised.

In summary, the Huge-LQG presents an interesting potential challenge to the
assumption of homogeneity in the cosmological principle. Its proximity to the
CCLQG at the same redshift adds to that challenge. Switching attention from
galaxies in the relatively local universe to LQGs at redshifts $z \sim 1$ may
well have advantages for such testing since the broad features of the
structures can be seen with some clarity, although, of course, the fine
details cannot.

\section{Acknowledgments}

LEC received partial support from the Center of Excellence in Astrophysics
and Associated Technologies (PFB 06), and from a CONICYT Anillo project (ACT
1122).

SR is in receipt of a CONICYT PhD studentship.


The referee, Maret Einasto, is thanked for helpful comments.

This research has used the SDSS DR7QSO catalogue \citep{Schneider2010}.

Funding for the SDSS and SDSS-II has been provided by the Alfred P. Sloan
Foundation, the Participating Institutions, the National Science
Foundation, the U.S. Department of Energy, the National Aeronautics and
Space Administration, the Japanese Monbukagakusho, the Max Planck
Society, and the Higher Education Funding Council for England. The SDSS
Web Site is http://www.sdss.org/.

The SDSS is managed by the Astrophysical Research Consortium for the
Participating Institutions. The Participating Institutions are the American
Museum of Natural History, Astrophysical Institute Potsdam, University of
Basel, University of Cambridge, Case Western Reserve University, University
of Chicago, Drexel University, Fermilab, the Institute for Advanced Study,
the Japan Participation Group, Johns Hopkins University, the Joint Institute
for Nuclear Astrophysics, the Kavli Institute for Particle Astrophysics and
Cosmology, the Korean Scientist Group, the Chinese Academy of Sciences
(LAMOST), Los Alamos National Laboratory, the Max-Planck-Institute for
Astronomy (MPIA), the Max-Planck-Institute for Astrophysics (MPA), New Mexico
State University, Ohio State University, University of Pittsburgh, University
of Portsmouth, Princeton University, the United States Naval Observatory, and
the University of Washington.

\bsp

\label{lastpage}


\begin{thebibliography}{}
%
%
  \bibitem[\protect\citeauthoryear{Brand et al.}{2003}]{Brand2003}
    Brand K., Rawlings S., Hill G.J., Lacy M., Mitchell E., Tufts J., 2003,
    MNRAS, 344, 283
%
  \bibitem[\protect\citeauthoryear{Buchert}{2011}]{Buchert2011}
    Buchert T., 2011, CQG, 28, 164007
%
  \bibitem[\protect\citeauthoryear{Clowes \& Campusano}{1991}]{Clowes1991}
    Clowes R.G., Campusano L.E., 1991, MNRAS, 249, 218
%
  \bibitem[\protect\citeauthoryear{Clowes et al.}{2012}]{Clowes2012}
    Clowes R.G., Campusano L.E., Graham M.J., S\"ochting I.K., 2012, MNRAS,
    419, 556
%
  \bibitem[\protect\citeauthoryear{Copi et al.}{2010}]{Copi2010}
    Copi C.J., Huterer D., Schwarz D.J., Starkam G.D., 2010, Adv. Astr.,
    2010, 847541
%
  \bibitem[\protect\citeauthoryear{Crampton, Cowley \& Hartwick}
  {Crampton et al.}{1987}]{Crampton1987}
    Crampton D., Cowley A.P., Hartwick F.D.A., 1987, ApJ, 314, 129
%
  \bibitem[\protect\citeauthoryear{Crampton, Cowley \& Hartwick}
  {Crampton et al.}{1989}]{Crampton1989}
    Crampton D., Cowley A.P., Hartwick F.D.A., 1989, ApJ, 345, 59
%
  \bibitem[\protect\citeauthoryear{Einasto et al.}{2011a}]{Einasto2011a}
    Einasto J. et al., 2011a, A\&A, 534, A128
%
  \bibitem[\protect\citeauthoryear{Einasto et al.}{2011b}]{Einasto2011b}
    Einasto M., Liivam\"agi L.J., Tago E., Saar E., Tempel E.,
    Einasto J., Mart\'inez V.J., Hein\"am\"aki P., 2011b, A\&A, 532, A5
%
  \bibitem[\protect\citeauthoryear{Einasto et al.}{2011c}]{Einasto2011c}
    Einasto M. et al., 2011c, ApJ, 736, 51
%
  \bibitem[\protect\citeauthoryear{Feldman, Watkins \& Hudson}
    {Feldman et al.}{2010}]{Feldman2010}
    Feldman H.A., Watkins R., Hudson M.J., 2010, MNRAS, 407, 2328
%
  \bibitem[\protect\citeauthoryear{Frith et al.}{2003}]{Frith2003}
    Frith W.J., Busswell G.S., Fong R., Metcalfe N., Shanks T., 2003,
    MNRAS, 345, 1049
%
  \bibitem[\protect\citeauthoryear{Geller \& Huchra}{1989}]{Geller1989}
    Geller M.J., Huchra J.P., 1989, Sci, 246, 897
%
  \bibitem[\protect\citeauthoryear{Gott et al.}{2005}]{Gott2005}
    Gott J.R., III, Juri\'c M., Schlegel D., Hoyle F., Vogeley M.,
    Tegmark M., Bahcall N., Brinkmann J., 2005, ApJ, 624, 463
%
  \bibitem[\protect\citeauthoryear{Hoyle et al.}{2012}]{Hoyle2012}
    Hoyle B., Tojeiro R., Jimenez R., Heavens A., Clarkson C., Maartens R.,
    2012, astro-ph/1209.6181
%
  \bibitem[\protect\citeauthoryear{\hbox{Hutsem\'ekers et al.}}{2005}]{Hutsemekers2005}
    Hutsem\'ekers D., Cabanac R., Lamy H., Sluse D., 2005, A\&A, 441, 915
%
  \bibitem[\protect\citeauthoryear{Jackson}{2012}]{Jackson2012}
    Jackson J.C., 2012, MNRAS, 426, 779
%
  \bibitem[\protect\citeauthoryear{Kashlinsky et al.}{2010}]{Kashlinsky2010}
    Kashlinsky A., Atrio-Barandela F., Ebeling H., Edge A., Kocevski D.,
    2010, ApJ, 712, L81
%
  \bibitem[\protect\citeauthoryear{Kashlinsky et al.}{2008}]{Kashlinsky2008}
    Kashlinsky A., Atrio-Barandela F., Kocevski D., Ebeling H., 
    2008, ApJ, 686, L49
%
  \bibitem[\protect\citeauthoryear{Komberg, Kravtsov \& Lukash}
    {Komberg et al.}{1996}]{Komberg1996}
    Komberg B.V., Kravtsov A.V., Lukash V.N., 1996, MNRAS, 282, 713
%
  \bibitem[\protect\citeauthoryear{Kubo et al.}{2007}]{Kubo2007}
    Kubo J.M., Stebbins A., Annis J., Dell'Antonio I.P., Lin H.,
    Khiabanian H., Frieman J.A., 2007, ApJ, 671, 1466
%
  \bibitem[\protect\citeauthoryear{Lietzen et al.}{2009}]{Lietzen2009}
    Lietzen H. et al., 2009, A\&A, 501, 145
%
  \bibitem[\protect\citeauthoryear{\hbox{Liivam\"agi}, Tempel \& Saar}
    {Liivam\"agi et al.}{2012}]{Liivamagi2012}
    Liivam\"agi L.J., Tempel E., Saar E., 2012, A\&A, 539, A80
%
  \bibitem[\protect\citeauthoryear{Longo}{1991}]{Longo1991}
    Longo M.J., 1991, ApJ, 372, L59
%
  \bibitem[\protect\citeauthoryear{Luparello et al.}{2011}]{Luparello2011}
    Luparello H., Lares M., Lambas D.G., Padilla N., 2011, MNRAS, 415, 964
%
  \bibitem[\protect\citeauthoryear{Marinoni, Bel \& Buzzi}
  {Marinoni et al.}{2012}]{Marinoni2012}
    Marinoni C., Bel. J., Buzzi A., 2012, JCAP, 10, 036
%
  \bibitem[\protect\citeauthoryear{Miller et al.}{2004}]{Miller2004}
    Miller L., Croom S.M., Boyle B.J., Loaring N.S., Smith R.J., Shanks T.,
    Outram P., 2004, MNRAS, 355, 385
%
  \bibitem[\protect\citeauthoryear{Nabokov \& Baryshev}{2008}]{Nabokov2008}
    Nabokov N.V., Baryshev Yu.V., 2008, in Baryshev Yu., Taganov I.,
    Teerikorpi P., eds, Practical Cosmology, Proceedings of the International
    Conference ``Problems of Practical Cosmology'', TIN, St. Petersburg,
    p. 69 (astro-ph/0809.2390)
%
  \bibitem[\protect\citeauthoryear{Padmanabhan et al.}{2007}]{Padmanabhan2007}
    Padmanabhan N. et al., 2007, MNRAS, 378, 852
%
  \bibitem[\protect\citeauthoryear{Park et al.}{2012}]{Park2012}
    Park C., Choi Y.-Y., Kim J., Gott J.R., III, Kim S.S., Kim K.-S., 
    2012, ApJ, 759, L7
%
  \bibitem[\protect\citeauthoryear{Pilipenko}{2007}]{Pilipenko2007}
    Pilipenko S.V., 2007, Astron. Rep., 51, 820
%
  \bibitem[\protect\citeauthoryear{Proust et al.}{2006}]{Proust2006}
    Proust D. et al., 2006, A\&A, 447, 133
%
  \bibitem[\protect\citeauthoryear{Raghunathan et al.}{2012}]{Raghunathan2012}
    Raghunathan S. et al., 2012, in preparation
%
  \bibitem[\protect\citeauthoryear{Richards et al.}{2002}]{Richards2002}
    Richards G.T. et al., 2002, AJ, 123, 2945
%
  \bibitem[\protect\citeauthoryear{Richards et al.}{2006}]{Richards2006}
    Richards G.T. et al., 2006, AJ, 131, 2766
%
  \bibitem[\protect\citeauthoryear{Rossmanith et al.}{2012}]{Rossmanith2012}
    Rossmanith G., Modest H., R\"ath C., Banday A.J., G\'orski K.M.,
    Morfill G., 2012, Phys. Rev. D 86, 083005
%
  \bibitem[\protect\citeauthoryear{Rozgacheva et al.}{2012}]{Rozgacheva2012}
    Rozgacheva I.K., Borisov A.A., Agapov A.A., Pozdneev I.A.,
    Shchetinina O.A., 2012, astro-ph/1201.5554
%
  \bibitem[\protect\citeauthoryear{Schneider et al.}{2010}]{Schneider2010}
    Schneider D.P. et al., 2010, AJ, 139, 2360
%
  \bibitem[\protect\citeauthoryear{Scrimgeour et al.}{2012}]{Scrimgeour2012}
    Scrimgeour M.I. et al., 2012, MNRAS, 425, 116
%
  \bibitem[\protect\citeauthoryear{Shen et al.}{2011}]{Shen2011}
    Shen Y. et al., 2011, ApJS, 194, 45 
%
  \bibitem[\protect\citeauthoryear{Sheth \& Diaferio}{2011}]{Sheth2011}
    Sheth R.K., Diaferio A., 2011, MNRAS, 417, 2938
%
  \bibitem[\protect\citeauthoryear{\hbox{S\"ochting}, Clowes \& Campusano}
  {S\"ochting et al.}{2002}]{Soechting2002}
    S\"ochting I.K., Clowes R.G., Campusano L.E., 2002, MNRAS, 331, 569
%
  \bibitem[\protect\citeauthoryear{\hbox{S\"ochting}, Clowes \& Campusano}
  {S\"ochting et al.}{2004}]{Soechting2004}
    S\"ochting I.K, Clowes R.G., Campusano L.E., 2004, MNRAS, 347, 1241
%
  \bibitem[\protect\citeauthoryear{Sylos Labini}{2011}]{SylosLabini2011}
    Sylos Labini F., 2011, EPL, 96, 59001
%
  \bibitem[\protect\citeauthoryear{Thomas, Abdalla \& Lahav}
  {Thomas et al.}{2011}]{Thomas2011}
    Thomas S.A., Abdalla F.B., Lahav O., 2011, Phys. Rev. Lett., 106, 241301
%
  \bibitem[\protect\citeauthoryear{Tsagas}{2012}]{Tsagas2012}
    Tsagas C.G., 2012, MNRAS, 426, L36
%
  \bibitem[\protect\citeauthoryear{Vanden Berk et al.}{2005}]{Vanden-Berk2005}
    Vanden Berk D.E., et al., 2005, AJ, 129, 2047
%
  \bibitem[\protect\citeauthoryear{Watkins, Feldman \& Hudson}
    {Watkins et al.}{2009}]{Watkins2009}
    Watkins R., Feldman H.A., Hudson M.J., 2009, MNRAS, 392, 743
%
  \bibitem[\protect\citeauthoryear{Webster}{1982}]{Webster1982}
    Webster A.S., 1982, MNRAS, 199, 683
%
  \bibitem[\protect\citeauthoryear{Williger et al.}{2002}]{Williger2002}
    Williger G.M., Campusano L.E., Clowes R.G., Graham M.J., 2002,
    ApJ, 578, 708
%
  \bibitem[\protect\citeauthoryear{Yadav, Bagla \& Khandai}
    {Yadav et al.}{2010}]{Yadav2010}
    Yadav J.K., Bagla J.S., Khandai N., 2010, MNRAS, 405, 2009
%
  \bibitem[\protect\citeauthoryear{Yershov, Orlov \& Raikov}
    {Yershov et al.}{2012}]{Yershov2012}
    Yershov V.N., Orlov V.V., Raikov A.A., 2012, MNRAS, 423, 2147
%
\end{thebibliography}
\end{document}